\input phyzzx
\nonstopmode
\sequentialequations
\twelvepoint
\nopubblock
\tolerance=5000
\overfullrule=0pt

\line{\hfill IASSNS-HEP 96/93}
\line{\hfill cond-mat/9609094}
\line{\hfill September 1996}
\titlepage
\title{Populated Domain Walls}
\vskip.2cm
 
\author{Chetan Nayak\foot{Research supported in part by a Fannie
and John Hertz Foundation fellowship.~~~
nayak@itp.ucsb.edu. Address as of Sept. 1, 1996:
Institute for Theoretical Physics, University of California,
Santa Barbara, CA 93106-4030}}
\vskip .2cm
\centerline{{\it Department of Physics }}
\centerline{{\it Joseph Henry Laboratories }}
\centerline{{\it Princeton University }}
\centerline{{\it Princeton, N.J. 08544 }}
 
\vskip.2cm
 
\author{Frank Wilczek\foot{Research supported in part by DOE grant
DE-FG02-90ER40542.~~~wilczek@sns.ias.edu}}
\vskip.2cm
\centerline{{\it School of Natural Sciences}}
\centerline{{\it Institute for Advanced Study}}
\centerline{{\it Olden Lane}}
\centerline{{\it Princeton, N.J. 08540}}
\endpage

\abstract{Several experiments suggest that the charge carriers
in the normal state of certain cuprate superconductors reside
on domain walls.  In an earlier paper, we suggested that several
aspects of the anomalous dynamical behavior of these materials
could be explained, at least qualitatively, on this basis.
Here, using results on the ground state energy of the 1-dimensional
Hubbard model (soluble by Bethe {\it ansatz\/} techniques) as a
function of charge density, we
argue that a non-zero charge density localized to
domain walls is a very plausible consequence
of strong short-range electron-electron repulsion.  We also suggest a method
to suppress meandering of the walls, and thereby enhance their signature
in neutron diffraction.}

\endpage

\REF\tranquada{J.M Tranquada {\it et al.}, Nature {\bf 375}, 561 (1995).}

\REF\birgenau{K. Yamada {\it et al.}, {\it Doping Dependence
of Spatially Modulated Dynamical Spin Correlations in Superconducting}
$~{\rm La}_{2-x}{\rm Sr}_x{\rm CuO}_4$, Tohoku Univ. preprint
(unpublished,
1996). }

\REF\poilblanc{D. Poilblanc and T.M. Rice, Phys. Rev. {\bf B 39}, 9749 (1989).}

\REF\schulz{H.J. Schulz, J Physique {\bf 50}, 2833 (1989).}

\REF\zaanen{J. Zaanen and O. Gunnarson, Phys. Rev. {\bf B 40}, 7391 (1989);
J. Zaanen {\it et al.}, Phys. Rev. {\bf B 53}, 8671 (1996).}

\REF\zzaanen{J. Zaanen and A. Oles, Ann. Physik, {\bf 5}, 224 (1996).}

\REF\emerykivelson{V.J. Emery and S.A. Kivelson, Physica {\bf C 209},
597 (1993); Physica {\bf C 235-240}, 189 (1994); in Strongly
Correlated Electronic Materials: The Los Alamos Symposium 1993,
eds. K.S. Bedell {\it et al.}, Addison-Wesley, Reading, MA 1994.}

\REF\castroneto{A.H. Castro-Neto and D.W. Hone, Phys. Rev. Lett. {\bf 76}, 
2165 (1996).}

\REF\nw{C. Nayak and F. Wilczek, Int. J. Mod. Phys {\bf B 10}, 2125 (1996).}

\REF\exclusion{F.D.M. Haldane, Phys. Rev. Lett. {\bf 67}, 937
(1991).}

\REF\emkivtop{S.A. Kivelson and V.J. Emery, cond-mat/9603009.}

\REF\liebwu{E.H. Lieb and F.Y. Wu, Phys. Rev. Lett. {\bf 20}, 1445
(1968).}

\REF\prel{P. Prelovsek, Phys. Rev. {\bf B47}, 5984 (1993); Phys. Rev.
{\bf B49}, 15421 (1994).}

\REF\scal{S. White and D. Scalapino, cond-mat/9610104.  This paper
appeared after the present one was completed and submitted.}

\REF\rice{F.C. Zhang and T.M. Rice, Phys. Rev. {\bf B 37},
3759 (1988).}

\REF\tnickel{J.M. Tranquada {\it et al.}, Phys. Rev. Lett. {\bf 73},
1003 (1994).}

\REF\zlitt{J. Zaanen and P. Littlewood, Phys. Rev. {\bf B50} 7222 (1994). } 

\REF\bell{S.N. Coppersmith {\it et al.}, Phys. Rev. {\bf B 25},
349 (1982).}

Recently there has been great interest in the electronic structure of
domain walls in effectively 2-dimensional materials with
antiferromagnetic order.  Some materials in this class have empty
walls, but there is mounting experimental evidence from neutron
scattering that at least in some of the cuprate materials the length
of the walls is a non-trivial multiple of the hole density
[\tranquada,\birgenau]; indeed, the electron filling fraction on the
wall appears to be approximately 1/4 -- one electron per every two
lattice sites -- in a variety of circumstances.  If correct this is a
very significant result, because it implies the possibility of charge
transport along the walls.  Effectively, the populated walls would
form a system of dynamical 1-dimensional wires within the
2-dimensional planes.

Even prior to the most recent experimental results the possibility
that spatially inhomogeneous structures occur in the cuprates, and are
important in understanding their markedly anomalous properties, was
the subject of considerable theoretical work
[\poilblanc,\schulz,\zaanen,\zzaanen,\emerykivelson,\castroneto].
In a recent paper [\nw ], written after
[\tranquada] but just before [\birgenau], we suggested
a particular hypothesis -- the minimal domain wall hypothesis --
which leads to domain walls having one hole per every two lattice
sites.    (It is interesting to note
that this hypothesis might be rephrased as the existence
of exclusion statistics [\exclusion] $g=2$ for spinless holons on the wall.)
Related ideas were suggested, independently, by Emery and Kivelson
under the title topological doping [\emkivtop]. 
We also suggested how several of the
most striking anomalous features of the normal state of the cuprates
could be understood, at least qualitatively, given the existence of
1-dimensional walls supporting non-trivial charge transport.

Unfortunately, however, detailed studies of wall energetics in the
Hartree-Fock approximation [\poilblanc,\schulz]
have invariably suggested that empty walls
are most favorable.
This puts the existence of the suggested universality
class open to doubt; more optimistically interpreted, it poses the
challenge to find an alternative, appropriate
approximation which makes it plausible
that populated walls can be favorable.

Here we shall present an energetic argument,
appropriate to electronic
systems with very strong short range repulsion (for which
Hartree-Fock is unreliable) which suggests physically interesting
conditions under which populated walls are energetically favored.
Recall that in a Hartree-Fock calculation, an expectation
value for the staggered magnetization is assumed, which
leads to the opening of a gap at the boundary of
the magnetic zone (which, in a model with only nearest-neighbor hopping, is the
Fermi surface at half filling). In the
presence of a domain wall, across which the staggered magnetization
changes sign, there are, in addition, states which lie in the gap.
Since their total length is smaller for a fixed number of holes,
empty domain walls cause less
disruption of the antiferromagnetic order
than do domain walls with non-zero filling.
As a result, they are energetically favored.
Here, we consider the opposite limit of very
large $U$ in which a one-electron picture is not
valid. The advantage of a domain wall with non-zero
filling is that electrons on the wall can gain
kinetic energy $t$ which is much greater than $J$
-- but still much smaller than $U$, of course, which prevents
them from escaping the wall. Said differently, in this limit,
the states on the wall lie within the lower Hubbard band
and therefore might be occupied. In a Hartree-Fock picture
on the other hand, they lie
between the lower and upper bands
which makes them energetically unfavorable.

Central to our argument is an old result of Lieb and Wu [\liebwu] on
the ground state energy of the one-dimensional Hubbard model.
The Hamiltonian of this classic
model contains two parameters $t$ and $U$, with the
units of energy, which parametrize the amplitude for hopping and
the on-site repulsion respectively.  The general result
is quite complicated,
but in the limit $U\rightarrow \infty$ takes the
following remarkably simple form:
$$
E_{\rm kinetic}/ N_s ~=~ -{t\sin 2\pi f} ~,
\eqn\kineticE
$$
where $f \equiv N_e/2N_s $ is the filling fraction, $N_e$ and
$N_s$ being respectively the number of electrons and the number
of sites along the wall.  Note that this energetic effect of
hopping remains finite as $U\rightarrow \infty$.  One can crudely
interpret the $U\rightarrow \infty $ result \kineticE\ as representing
the energy of free {\it spinless\/} fermions: the exclusion constraint
between different spins is satisfied, in the ground state, by
enforcing
complete antisymmetry.
Remarkably, this energy is indeed
minimized at $f={1\over 4}$, the value suggested by the minimal
domain wall hypothesis.

That is not by any means the full story, however.  We are really
more interested in the minimum energy per hole, not per wall
site.  Furthermore, the energy involved in creating the walls must
be taken into account.

Let us step back a moment to roughly
survey the problem of energetics for
holes doped into an antiferromagnetic background,
with large short-range
(say, for simplicity, on-site) repulsion.
If one sprinkles isolated
holes into such a background, simply by emptying sites, then each
hole removes 4 potential $J {{\bf S}_i}\cdot{{\bf S}_j}$
spin-spin alignment terms, so
that there is a penalty
$4\lambda J$ in energy. (Breaking a singlet bond costs
energy $-{3\over 4} J$. In an antiferromagnet, one cannot have
singlets throughout, 
the energy per bond is reduced, $\lambda < {3\over 4}$.)
One can cut this down by allowing the holes to
form connected patches; then the energy cost is $2\lambda J$ per hole.
Alternatively one can put the holes on walls; the cost of this
is $3\lambda J$ per wall site.  The question then becomes whether the
energy gained by allowing for hopping, as extracted above, can
allow the walls to compete successfully against the patches,
and if so, at what filling fractions.

The total energy per hole, if they are localized to walls at
electron filling fraction $f$, is
$$
E_{\rm T}/N_h ~=~ {{3\lambda J}\over 2 f^\prime } -
  {t\sin 2\pi f^\prime \over \pi f^\prime}~,
\eqn\totalE
$$
where $f^\prime \equiv {1\over 2} - f$ is the hole filling fraction.
Defining $r\equiv {3\pi\lambda J\over {2t}}$, and $x\equiv 2\pi f^\prime$,
we are led to minimize the function
$$
h(x) ~=~ {r\over x} - {\sin x \over x}
\eqn\cleanedup
$$
in the interval $0\leq x \leq \pi$.
A short analysis shows that
to insure favorability against the patch configuration we
must require $\sin x \geq {r\over 3}$.
For very small $r$ the minimum occurs at
$x \approx (3r)^{1\over 3}$, a nearly filled
wall; for $r=1$ it occurs at
$x = {\pi\over 2}$,
the `minimal domain wall' value;
for $r\geq \pi$, the minimum occurs at $x=\pi$, an empty wall.
Using Newton's method, one readily calculates that near this
value the filling fraction changes with $r$ as
$$
{\biggl({df\over dr}\biggr)_{|r=1}} ~=~  {1\over \pi^2}~.
\eqn\changeoff
$$
Populated domain walls
remains favorable compared to patches up to $r\approx 1.4$,
corresponding to $f^\prime \approx .43$.
Because of the simple form of the energy function $h$, there
is never any advantage to having coexisting walls and patches.

The calculation presented above is far from rigorous, primarily
because the antiferromagnetic background is regarded as given and
inert, while it properly should be regarded as composed of dynamical
electrons on the same footing as those in the walls.  It could
easily be formulated as a variational calculation, though again with
no firm control on the errors.  In its defense, we can fairly claim
the virtue of simplicity, and that the terms retained plausibly represent
basic effects having clear physical interpretations.
Taken at face value, its result certainly
suggests that the ground state
of an antiferromagnetic system with strong short-range repulsion
will generically contain
populated domain walls.

Within this framework there is as
yet no
clear sign that $f={1\over 4}$ is especially favored.  It is
worth noting, however, that for effective
spinless fermions, which as mentioned above we seem to have at least 
in some approximate sense, $f={1\over 4}$ corresponds to half filling;
and when phonon interactions are taken 
into account there is a favorable commensuration energy near this value,
due to the possibility of a Peierls distortion, as pointed out by
Zaanen and Oles [\zzaanen ].  One can
model
this by using a modified dispersion relation for the electrons; but we
leave
that (and other refinements) for future work.

A few other comments are in order. First: we have neglected the
effects of the long-range part of the Coulomb interaction.  As has
been emphasized by Emery and Kivelson [\emerykivelson], this
interaction strongly disfavors large clusters of holes; but walls with
non-zero electron filling are less affected [\tranquada ].  Second:
formally, one could both gain in hopping energy and break fewer bonds
by merging two (or more) populated walls.  Each wall site would then
break $2{1\over 2}$ sites instead of 3, and at 1/4 filling a short
calculation shows that one gains about 10\% in hopping energy per
electron, in so far as the electrons are effectively free spinless
fermions.  However one would pay a heavy price in Coulomb energy; one
sign of this is that the electrostatic force between separated walls
is certainly repulsive, so that a configuration with uniformly spaced
walls of unit thickness is at least locally stable.  Entropic
considerations also favor the separated walls.  Third: although we
have been using the term domain wall, at no point did we require the
staggered magnetization to change sign across the wall (unlike in a
Hartree-Fock calculation). If it did not change sign, however, there
would be an effective staggered magnetic field at the wall and
electrons on the wall would tend to form a spin-ordered state. The
principal effects of the interaction between the wall and the
antiferromagnetic regions which surround it would be a renormalization
of the hopping amplitude and fluctuation in the position of the wall
(see below), both effects are important quantitatively, but do not
affect our qualitative conclusion.  Finally, there is some numerical
evidence, from simulations of the $t-J$ model on small lattices, that
can be interpreted as indicating the existence of populated domain
walls [\prel, \scal ].  The latter study suggests that the walls two
lattice spacings thick might be most favorable, which is not
inconsistent with our general picture.  We plan to investigate these
more delicate aspects of the energetics analytically in the near
future; again, our main point here is that what is presumably the
dominant energetic effect, namely the short-range repulsion, seems
strongly to favor populated domain walls in some form.

It is interesting now to apply these considerations to
actual materials, using the
parameters of $t-J$ models as
extracted from phenomenological models fitted to their
low-energy behavior [\rice]. They find $J/t \approx 1/4$. 
Numerical estimates
of the energy per bond of a 2D Heisenberg antiferromagnet
give $\lambda \approx 0.5$.  Formally, these estimates
lead us to an
electron filling fraction $f=.29$ on the wall, which is
significantly (though not grossly)
larger
than $1/4$; clearly, however, our approximations have been
too crude to  
support close numerical comparison with experiment.  
Systems with substantially larger values of $J/t$, such as 
doped La$_2$NiO$_4$ studied experimentally by Tranquada {\it et al}.
[\tnickel ] and theoretically by Zaanen and Littlewood [\zlitt ], are
predicted to have small or vanishing wall occupation.

All this certainly encourages us to take the possibility that
populated domain walls, generally curved and fluctuating in space
and time, can be important dynamical objects.  A
direct
experimental signature for inhomogeneous
structures could come from neutron scattering,
but in practice it requires that these structures
be at least quasi-static and fairly regularly arrayed.  One would
expect that running currents through the walls (possible if they
contain the charge carriers!)  would tend to straighten them out
and suppress fluctuations, just as a garden hose straightens out
when water runs through it.

To make this intuition quantitative,
let us consider the energy governing the fluctuations.  Since it
is proportional to the number of broken antiferromagnetic bonds, it
is
of the form
$$
\eqalign{
E_{\rm curv.} ~&=~ \int dx \alpha \sqrt {1+({dy\over dx})^2 } \cr
               &=~ {\rm const.}~ + \int dx \alpha ({dy\over dx})^2~,\cr}
\eqn\curvener
$$
for a wall in the $x$ direction.
Here the tension $\alpha$ is
proportional to $J/a$, where $J$ is
the effective antiferromagnetic (super)exchange interaction, and
$a$ the lattice spacing; in the second equation we have assumed the
wall is nearly straight.   If we think of $x$ as a mock time variable,
then the wall executes a random walk in the $y$-direction; such
fluctuations will disorder an array of walls.  A very similar
problem was considered in the context of krypton adsorbed on graphite
in [\bell] and for antiferromagnetic domain walls (as here) in
[\zaanen].  The domain walls form a fluid
at high temperatures, but at low
temperatures undergo a Kosterlitz-Thouless transition to an ordered
phase.  The neutron scattering data alluded to above, interpreted
in this way, require that the structural phase transition stabilizes
the ordered phase in La$_{1.48}$Nd$_{0.4}$Sr$_{0.12}$CuO$_4$, while
La$_{2-x}$Sr$_x$CuO$_4$ is in the fluid phase.

Now let us consider the current-carrying state.  The voltage produces
an electrostatic energy of the form
$$
E_{\rm estat.} ~=~ \int dx \lambda Ex \sqrt {1+({dy\over dx})^2}~,
\eqn\exenerg
$$
where $\lambda$ is the charge density along the wall; while the force
exerted on a current $i$ by a wall that bends through angle
$\theta$ is $\hbar k_F \theta i/e$, corresponding to an energy functional
$$
E_{\rm current} ~=~ \int dx \hbar k_F i/e ({dy\over dx})^2 ~.
\eqn\ienerg
$$
The latter contribution the energy dominates and will be significant when
it becomes comparable to the temperature.  An order-of-magnitude
calculation indicates that a current per domain wall of
10$^{-7}$ A, corresponding to a current density
$\approx 10^7 {\rm {A} \over \rm {cm.}^2}$, which is very large but
perhaps not prohibitive.

Finally, we note that by applying pressure it should be possible to
vary
the effective $t$ and thereby, modify the favored wall filling factor.

\bigskip
\bigskip

{\bf Acknowledgment}:  C.N. would like to thank A.H. Castro-Neto and
M.P.A. Fisher for discussions.

\endpage

\refout

\endpage

\end